\documentclass[epj,twocolumns]{svjour_mod}
\usepackage{amsmath} 
\usepackage{amssymb}
\usepackage{graphicx,epsfig}
\usepackage[mathscr]{eucal}
\usepackage{color}
\usepackage{longtable}
\usepackage[english]{babel}
\usepackage{tabularx}
\usepackage{setspace}
\usepackage{xspace}
\usepackage{multirow}
\usepackage{cite}
\renewcommand{\thefigure}{\arabic{figure}}

\newcommand{\gev}{\mathrm{GeV}}
\newcommand{\mw}{m_W}

\newcommand{\ptlep}{p_\perp^\ell}
\newcommand{\ptlepmin}{p_\perp^{\ell,\rm min}}
\newcommand{\ptlepmid}{p_\perp^{\ell,\rm mid}}
\newcommand{\ptlepmax}{p_\perp^{\ell,\rm max}}
\newcommand{\apt}{{\cal A}_{\ptlep}}

\newcommand{\MT}{M_\perp^{\ell\nu}}

\newcommand{\muq}{\mu_Q}

\newcommand{\ptz}{p_\perp^{\ell^+\ell^-}}
\newcommand{\ptw}{p_\perp^{\ell\nu}}

\newcommand{\cmw}{{\cal C}_{m_W}}

\usepackage{scalefnt,pstricks}

\usepackage{array}
\newcolumntype{L}[1]{>{\raggedright\let\newline\\\arraybackslash\hspace{0pt}}m{#1}}
\newcolumntype{C}[1]{>{\centering\let\newline\\\arraybackslash\hspace{0pt}}m{#1}}
\newcolumntype{R}[1]{>{\raggedleft\let\newline\\\arraybackslash\hspace{0pt}}m{#1}}

\usepackage[colorlinks=true,allcolors={blue!70!black}]{hyperref}

\begin{document} 

\author{Luca Rottoli \inst{1}
\and Paolo Torrielli \inst{2}
\and Alessandro Vicini \inst{3}
}

\date{}

\title{Determination of the $W$-boson mass at hadron colliders}

\institute{ Physik Institut, Universit\"at Z\"urich, CH-8057 Z\"urich, Switzerland
 \and Dipartimento di Fisica, Universit\`a di Torino and INFN, Sezione di Torino, I-10125 Torino, Italy
 \and Dipartimento di Fisica,
Universit\`a di Milano and INFN, Sezione di Milano, I-20133 Milano, Italy
}


\abstract{
We introduce an observable relevant for the determination of the $W$-boson mass $m_W$ at hadron colliders. This observable is defined as an asymmetry around the jacobian peak of the charged-lepton transverse-momentum distribution in the charged-current Drell-Yan process.
We discuss the observable's theoretical prediction, presenting results at different orders in QCD, and showing its perturbative stability.
Its definition as a single scalar number and its linear sensitivity to $m_W$ allow a clean extraction of the latter and a straightforward discussion of the associated theoretical systematics: a perturbative QCD uncertainty of ${\cal O}(\pm 5)$ MeV on $\mw$ can be established by means of this observable, relying solely on charged-current Drell-Yan information. Owing to its relatively inclusive nature, the observable displays desirable properties also from the experimental viewpoint, especially for the unfolding of detector effects. We show that a measurement of this observable can lead to a competitive experimental error on $m_W$ at the LHC.
}
\maketitle

\paragraph{Introduction.}

The experimental determination of the $W$-boson mass $\mw$ \cite{CDF:2013dpa,Aaboud:2017svj,LHCb:2021bjt,CDF:2022hxs} plays a central role in the programme of precision tests of the Standard Model (SM) at hadron colliders.
A potential discrepancy between the measured value and precise $\mw$ predictions \cite{Awramik:2003rn,Degrassi:2014sxa}
within the SM may immediately hint at the presence of New-Physics effects, as comprehensively discussed in the context of global fits \cite{Baak:2014ora,deBlas:2021wap} of electroweak (EW) precision observables.

At hadron colliders, the value of $\mw$ is primarily inferred from the analysis of the charged-current Drell-Yan (CCDY) process. Of particular relevance are the properties of final-state kinematical distributions defined in the transverse plane with respect to the collision axis, such as the charged-lepton transverse momentum $\ptlep$, the lepton-pair transverse mass $\MT$ and transverse momentum $\ptw$, and the missing transverse energy $E_T$~\cite{CDF:2013dpa,Aaboud:2017svj,LHCb:2021bjt,CDF:2022hxs}.

Experimental analyses aiming at the measurement of $\mw$ typically employ a QCD modelling of CCDY based on parton-shower Monte Carlo (MC) event generators, whose parameters are tuned on high-precision neutral-current Drell-Yan (NCDY) measurements, chiefly the lepton-pair transverse momentum $\ptz$.
A data-driven tuning step is in general necessary, as a standalone prediction of CCDY with the relatively low accuracy provided by MC simulations typically leads to an insufficient description of data.
Tuned MC predictions are then used to prepare templates of the relevant transverse kinematical distributions with different $\mw$ hypotheses. Theoretical templates are subsequently compared with CCDY experimental data, and a $\chi^2$ analysis is performed to determine the preferred value for $\mw$.

Such a significant dependence of the $\chi^2$-based approach on the tuning to NCDY experimental data poses however some conceptual issues for $\mw$ determination. On the one hand, the fit procedure heavily relies on phenomenological models rather than on first-principle SM predictions. In turn, this exposes the procedure to the risk of hiding New-Physics effects in the fit of model parameters. On the other hand, and even more severely, it
hinders the possibility to perform meaningful studies of the perturbative uncertainty associated with the theoretical prediction: even an MC tool with arbitrarily low formal accuracy can indeed yield an excellent description of data, provided it grants sufficient flexibility for tuning.
This approach essentially makes no use of the high-quality theoretical understanding of NCDY and CCDY lepton-pair production~\cite{Alioli:2016fum}, which in recent years has witnessed a substantial progress in the description of fixed-order~\cite{Duhr:2020sdp,Duhr:2020seh,Duhr:2021vwj,Chen:2021vtu,Boughezal:2015dva,Gehrmann-DeRidder:2015wbt,Boughezal:2015ded,Boughezal:2016dtm,Boughezal:2016isb,Gehrmann-DeRidder:2016cdi,Gehrmann-DeRidder:2016jns,Gauld:2017tww,Gehrmann-DeRidder:2017mvr,Gauld:2021pkr,Alekhin:2021xcu,Buonocore:2021tke,Camarda:2021jsw,Chen:2022lwc}
and all-order~\cite{Bizon:2018foh,Becher:2019bnm,Bizon:2019zgf,Ebert:2020dfc,Becher:2020ugp,Camarda:2021ict,Re:2021con,Ju:2021lah,Chen:2022cgv,Chen:2022lpw,Neumann:2022lft,Isaacson:2022rts} QCD effects, as well as in the evaluation of EW~\cite{Dittmaier:2001ay,Baur:2001ze,Baur:2004ig,Arbuzov:2005dd,Zykunov:2005tc,Zykunov:2006yb,CarloniCalame:2006zq,CarloniCalame:2007cd,Arbuzov:2007db,Dittmaier:2009cr} and mixed QCD-EW~\cite{Balossini:2008cs,Balossini:2009sa,Barze:2012tt,Barze:2013fru,Dittmaier:2014qza,Dittmaier:2015rxo,Bonciani:2016wya,deFlorian:2018wcj,Bonciani:2019nuy,Delto:2019ewv,Cieri:2020ikq,Bonciani:2020tvf,Buccioni:2020cfi,Behring:2020cqi,Buonocore:2021rxx,Bonciani:2021iis,Bonciani:2021zzf,Behring:2021adr,Buccioni:2022kgy} corrections.

Theoretical systematics in the data-driven procedure are mainly assessed by quantifying to what extent the experimental input from $\ptz$\!\! may be applied to $\ptw$, given the theoretical knowledge of the two distributions \cite{Giele:1998gw}, with limited further constraints coming from the direct measurement of $\ptw$ \cite{CDF:2022hxs,ATLAS:2023llf}: this might underestimate uncertainties, as it assumes that the procedure works equally well for all observables used for $\mw$ extraction.
The impact of modelling uncertainties on $\mw$ determination has been discussed considering
the role of parton distribution functions (PDFs) \cite{Bozzi:2011ww,Bozzi:2015hha,Bozzi:2015zja,Quackenbush:2015yra,Kotwal:2018qzl,Hussein:2019kqx,Bagnaschi:2019mzi,Gao:2022wxk},
of non-perturbative contributions to transverse spectra \cite{Bacchetta:2018lna}, as well as of EW and of leading QCD-EW corrections \cite{Dittmaier:2015rxo,CarloniCalame:2016ouw,Behring:2021adr};
all of these studies assume the existence of an underlying perturbative description of the process supplemented by a data-driven non-perturbative model. However, the quoted theoretical uncertainties typically neglect the interplay of the perturbative and the non-perturbative components.

In this letter we present an alternative strategy to determine the value of $\mw$ which fully exploits the theoretical progress in the description of Drell-Yan lepton-pair production.
We introduce a new observable based on the charged-lepton transverse-momentum distribution in CCDY, defined as an asymmetry around its jacobian peak at $\mw/2$. On the one hand, its clean definition in terms of calculable fiducial rates allows to directly interpret the extracted $\mw$ as the fundamental SM parameter; on the other hand, the observable displays excellent perturbative convergence,  which enables a robust study of the associated perturbative-QCD (pQCD) uncertainties, and its theoretical description is systematically improvable by adding subleading QCD and EW effects.
The simple dependence of the observable upon $\mw$ in turn allows a plain study of the impact of non-perturbative QCD (npQCD) effects, as well as a consistent propagation of their uncertainties in the prediction.

\paragraph{Lepton transverse momentum and sensitivity to $\mw$.}
\label{sec:ptlep}
The modelling of $\ptlep$ in CCDY
requires a precise description of the QCD contributions to the transverse and longitudinal degrees of freedom of the final state~\cite{Manca:2017xym}.
At leading order (LO) the charged lepton and the neutrino are back-to-back, $\ptw=0$, thus, neglecting lepton masses and the $W$-boson decay width $\Gamma_W$, the $\ptlep$ distribution has a sharp kinematical endpoint at $\ptlep=\mw/2$, which is the origin of its sensitivity to the $W$-boson mass (see also \cite{Rujula:2011qn,Bianchini:2019iey}).
Beyond LO in QCD, the region around the endpoint develops a sensitivity to soft radiation, which in turn generates an integrable singularity~\cite{Catani:1997xc} in the fixed-order differential $\ptlep$ spectrum.
The all-order treatment of soft and collinear initial-state QCD radiation, achieved by a resummation of enhanced logarithms $\log(\ptw/\mw)$, is therefore a central ingredient for a reliable description of $\ptlep$.
Such a resummation nowadays reaches next-to-next-to-next-to-leading-logarithmic (N$^3$LL) accuracy, matched with the next-to-next-to-leading-order (NNLO) predictions for the transverse-momentum spectrum~\cite{Bizon:2019zgf}.

\begin{figure}[t]
\begin{center}
\includegraphics[width=\columnwidth]{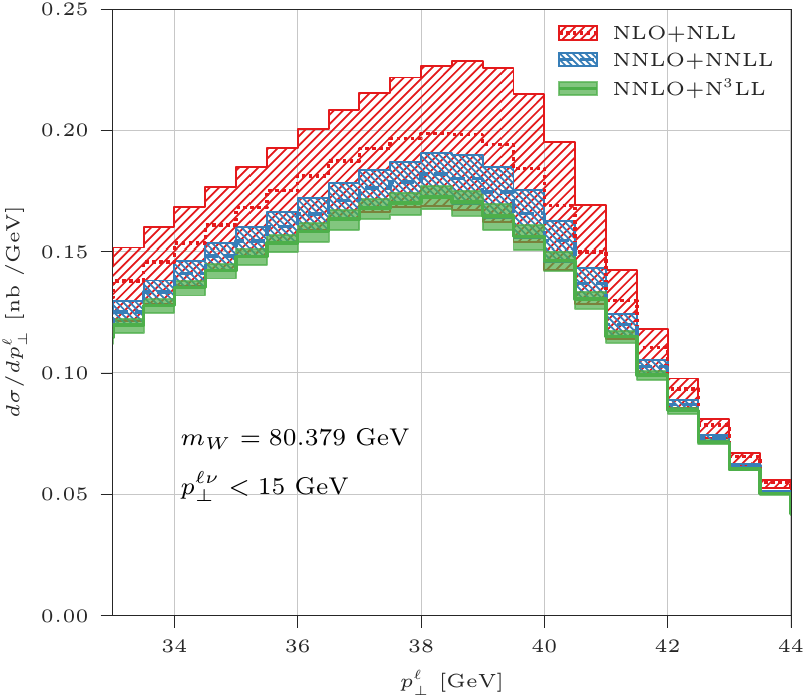}
\includegraphics[width=\columnwidth]{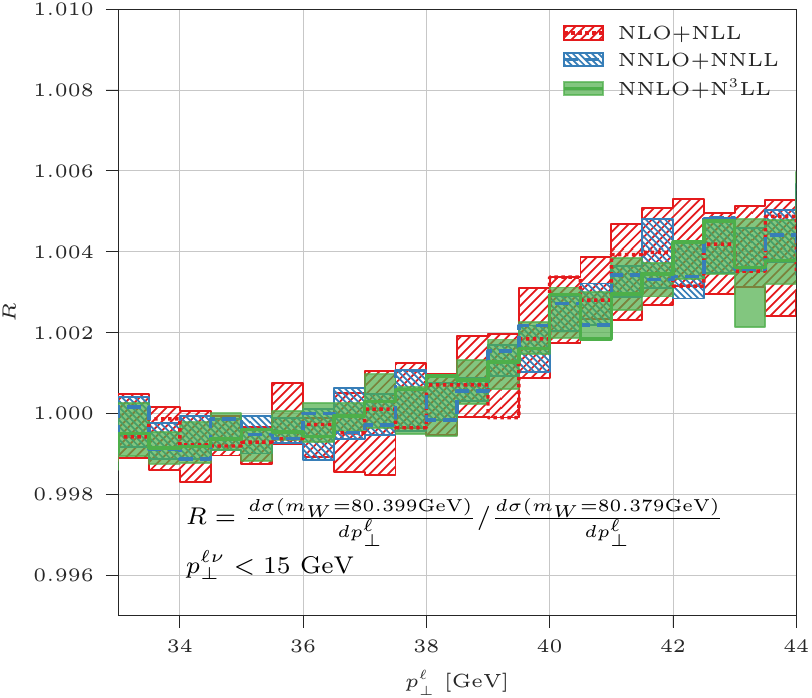}\\[2ex]
\end{center}
\vspace{-2ex}
\caption{\label{fig:ptlep}
Upper panel: charged-lepton transverse-momentum distribution in CCDY, computed with different QCD approximations and reference $\mw=80.379$ GeV. Lower panel: ratio of $\ptlep$ distributions computed with two $\mw$ values differing by 20 MeV. Uncertainty bands are obtained as the envelope of 9 renormalisation, factorisation and resummation scales, see text for further details.
}
\end{figure}
In the following, we consider the $\ptlep$ distribution in $W^-$ production at the Large Hadron Collider (LHC) with centre-of-mass energy $\sqrt{S}=13$ TeV and acceptance cuts
$\ptlep>20$ GeV,\,
$\MT > 27$ GeV,\,
$|\eta_\ell| <2.5$,\, 66 GeV $< M^{\ell\nu} < 116$ GeV, $\ptw < 15$ GeV \footnote{A $\ptw<15$ GeV cut, achievable in a Tevatron setup, is slightly lower than what currently employed by the LHC experimental collaborations. We stress however that all of the features discussed in the following remain qualitatively unchanged when applying a $\ptw < 30$ GeV cut.} ($\eta_\ell$ and $M^{\ell\nu}$ being the charged-lepton rapidity and the lepton-pair invariant mass, respectively), using the central replica of the NNPDF4.0 NNLO proton PDF set~\cite{NNPDF:2021njg} with strong coupling constant $\alpha_s(m_Z) = 0.118$ through the LHAPDF interface~\cite{Buckley:2014ana}.
We give predictions for three different QCD approximations, NLO+NLL, NNLO+NNLL and NNLO+N$^3$LL, using the \texttt{RadISH}~\cite{Monni:2016ktx,Bizon:2017rah,Monni:2019yyr,Re:2021con} code for $\ptw$ resummation, with a fixed-order prediction provided by \texttt{MCFM} \cite{Campbell:2019dru}.\sloppy
Here and in the following, the labels N$^k$LO (N$^k$LL) refer to the accuracy of the underlying CCDY cross section (of the resummed $\ptw$ spectrum).
We match the two results using the $q_T$-subtraction formalism~\cite{Catani:2007vq}, with a technical slicing cutoff $q_T^{\rm cut}=0.81$~GeV in the \texttt{MCFM} calculation.
Linear fiducial power corrections are included to all orders in the \texttt{RadISH} prediction through transverse recoil~\cite{Catani:2015vma,Ebert:2020dfc}.
We consider 21 values of $\mw$ between 80.329 GeV and 80.429 GeV, in steps of 5 MeV.
Renormalisation, factorisation and resummation scales are chosen as $\mu_{R,F}=\xi_{R,F}\sqrt{(M^{\ell\nu})^2 + (\ptw)^2}$, and $\muq=\xi_Q\,M^{\ell\nu}$, respectively. We estimate pQCD uncertainties by varying $\xi_R$ and $\xi_F$ independently in the range (1/2,\,1,\,2), excluding $\xi_{R,F}/\xi_{F,R} =4$, while keeping $\xi_Q=1/2$ (7 variations). In addition, we consider the 2 variations of $\xi_Q$ in (1/4,\,1) at central values $\xi_R = \xi_F = 1$, thereby obtaining a total envelope of 9 variations.

The upper panel of Figure~\ref{fig:ptlep} displays the perturbative convergence of the $\ptlep$ distribution, for a given value of $\mw=80.379$ GeV: one can notice how the inclusion of higher-order pQCD effects in resummed predictions
translates into a significant reduction of theoretical systematics.
The lower panel of Figure~\ref{fig:ptlep} shows with a ratio plot the impact on the $\ptlep$ distribution of a 20-MeV shift of the reference $\mw$ value. Such a shift induces a shape distortion at the $0.5\%$-level around the jacobian peak, an effect which is clearly resolvable beyond the theoretical uncertainty, assuming full correlation between the scales of numerator and denominator in the ratio. We also note that, starting from a baseline featuring all-order QCD radiation, the effect of the $\mw$ shift is remarkably independent of the QCD perturbative order and scale choice, as a consequence of the factorisation of initial-state QCD radiation from $W$-boson production and decay.

The sensitivity to $\mw$ of the $N$ bins $\sigma_i$ of the $\ptlep$ distribution can be quantified by means of the covariance matrix with respect to $\mw$ variations,
$\left(\cmw \right)_{ij} \equiv \langle\sigma_i \, \sigma_j\rangle - \langle\sigma_i\rangle \, \langle\sigma_j\rangle$, where the $\langle\,\rangle$ symbol indicates an arithmetic average over the different available $\mw$ options (21 in our case).
The $N$ eigenvectors of $\cmw$ represent the linear combinations of $\ptlep$ bins that transform independently under $\mw$ variations, and the corresponding eigenvalues
in turn express the sensitivity of such combinations to $\mw$.

For $\ptlep$ bins around the jacobian peak, such as those contributing to Figure~\ref{fig:ptlep}, there is a strong hierarchy among the $\cmw$ eigenvalues, with the first one being more than an order of magnitude larger than all others. Such a feature, robust against variations of the considered $\ptlep$ range, suggests the first linear combination to be representative of the behaviour of the whole $\ptlep$ distribution under $\mw$ variations.
In our simulation setup, the coefficients of this linear combination are all positive (negative) for bins at $\ptlep<37$ GeV ($\ptlep>37$ GeV), irrespectively of the employed QCD approximation or of the $\ptlep$ range.
The pattern of signs is in turn indicative of $\mw$ sensitivity: the value of 37 GeV is directly related to the position of the jacobian peak at $\mw/2$, after considering the smearing due to all-order QCD radiation as well as to the $W$-boson decay width (we set $\Gamma_W=2.084$ GeV).
Inspection of the lower panel of Figure~\ref{fig:ptlep} confirms the value $\ptlep=37$ GeV as separating the spectrum
into two regions, respectively with ($\ptlep>37$ GeV) and without ($\ptlep<37$ GeV) sensitivity to $\mw$.

\begin{figure}[t]
  \begin{center}
\includegraphics[width=\columnwidth]{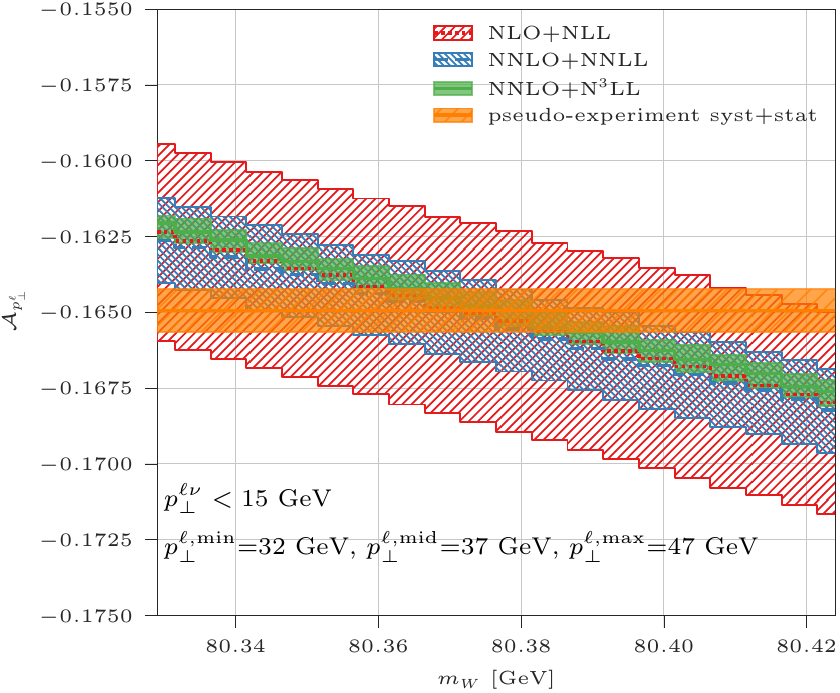}
\end{center}
\vspace{-2ex}
\caption{\label{fig:asymmetry}
The asymmetry $\apt$ as a function of $\mw$, in different QCD approximations.}
\end{figure}

\paragraph{Jacobian asymmetry and $\mw$ determination.}
\label{sec:asy}
Based on the previous considerations, we introduce a $\ptlep$ range $[\ptlepmin,\ptlepmax]$ which includes the jacobian peak, as well as an intermediate value $\ptlepmin < \ptlepmid < \ptlepmax$, and define two fiducial cross sections,
\begin{equation}
L_{\ptlep}
\equiv
\int_{\ptlepmin}^{\ptlepmid} d\ptlep \frac{d\sigma}{d\ptlep}
\, ,
\quad\,\,
U_{\ptlep}
\equiv
\int_{\ptlepmid}^{\ptlepmax} d\ptlep \frac{d\sigma}{d\ptlep}
\, ,
\label{eq:PDptlep}
\end{equation}
together with their asymmetry
\begin{equation}
\apt(\ptlepmin,\ptlepmid,\ptlepmax)
\, \equiv \,
\frac{L_{\ptlep}-U_{\ptlep}}
{L_{\ptlep}+U_{\ptlep}}
\, .
\label{eq:Aptlep}
\end{equation}
In Figure~\ref{fig:asymmetry} we plot $\apt(32\, \gev,37\, \gev,47\, \gev)$ as a function of $\mw$, with different QCD approximations.
The uncertainty bands computed (with the same scale choice for $L_{\ptlep}$ and $U_{\ptlep}$) at the various perturbative orders exhibit an excellent convergence pattern, and in all cases encompass predictions at the next orders.
Given this behaviour, we consider the size of the NNLO+N$^3$LL uncertainty band as a good estimator of the
uncertainty due to missing pQCD higher-order effects.
We have studied the dependence of this pattern on $\ptlepmid$ and found that for $\ptlepmid\gtrsim 38$ GeV, approaching the effective endpoint of the fixed-order distribution, the perturbative convergence slightly deteriorates; on the contrary, choices with $\ptlepmid < 37$ GeV exhibit a better stability, at the price of a reduced sensitivity to $\mw$. We then choose $\ptlepmid=37$ GeV as our default, as an excellent compromise between stability and sensitivity. The convergence behaviour is instead fairly stable against variations of $\ptlepmin$ and $\ptlepmax$.

We remark in Figure~\ref{fig:asymmetry} that $\apt$ has a clear linear sensitivity to $\mw$, directly stemming from the linear $\mw$-dependence of the jacobian-peak position. Moreover, its slope is extremely stable irrespectively of the QCD approximation and the scale choice, and just depends on the defining $\ptlep$ range, which reflects the factorisation of QCD initial-state radiation from the $\mw$-sensitive propagation and decay. These features make $\apt$ an excellent observable to determine $\mw$ and to robustly quantify the associated uncertainties.
For a given choice of $[\ptlepmin,\ptlepmid,\ptlepmax]$,
the experimental value of $\apt$ can be obtained by simply measuring the fiducial cross sections $L_{\ptlep},\,U_{\ptlep}$ (i.e.~a counting experiment), eventually resulting in a single scalar number in which systematic uncertainties can be straightforwardly propagated.
The relatively large size of the $[\ptlepmin,\ptlepmid]$ and $[\ptlepmid,\ptlepmax]$ intervals helps taming the statistical error, and would be beneficial with a view to unfolding detector effects, for a comparison with theory predictions at particle level; the latter is welcome in view of a combination of the results obtained by different experiments \cite{Amoroso:2022rly}.
For illustrative purposes, in Figure~\ref{fig:asymmetry} we plot a hypothetical experimental measurement for $\apt$, with statistical and systematic errors realistically propagated.

From Figure \ref{fig:asymmetry} we compare the experimental error band with a single theoretical curve (arbitrarily chosen, as all have the same slope): the intercepts of the curve with the edges of the band identify an $\mw$ interval that we treat as the experimental uncertainty.
The large CCDY cross section implies high statistical precision on $\apt$, and already with a luminosity of ${\cal L}=140$ fb$^{-1}$ we find $\Delta \apt^{\rm stat} = \pm \, 0.00007$; moreover, assuming a relative systematic error of 0.001 in the measurement of both $L_{\ptlep}$ and $U_{\ptlep}$, and neglecting experimental correlations in the error propagation, we obtain $\Delta \apt^{\rm syst} = \pm 0.0007$. Such numbers translate into an $\mw$ uncertainty $\Delta\mw^{\rm stat}+\Delta\mw^{\rm syst}\sim \pm \, 1.3 \, \pm \, 12.5$ MeV.
We then take the two edges of the scale-variation band at a given perturbative accuracy, and use them to estimate the uncertainty on $\mw$ due to missing pQCD higher orders, by comparison with the central experimental result. At NNLO+N$^3$LL we find a very competitive $\Delta\mw^{\rm pQCD}\sim \pm 6$ MeV.

\begin{figure}[t]
\begin{center}
\includegraphics[width=\columnwidth]{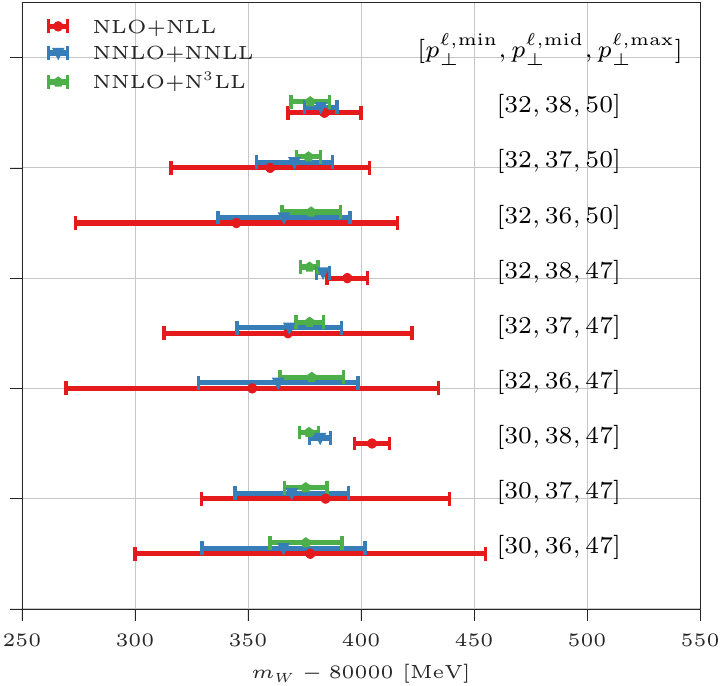}
\end{center}
\vspace{-2ex}
\caption{\label{fig:shifts}
The range of $\mw$ values obtained comparing
  the band of theoretical predictions at different orders in pQCD, with the central
  experimental value of $\apt$. Different choices of $[\ptlepmin,\ptlepmid,\ptlepmax]$
  are considered.
}
\end{figure}

In Figure~\ref{fig:shifts} we quantify the pQCD uncertainty on $\mw$ as just outlined, considering different perturbative orders and choices of $[\ptlepmin,\ptlepmid,\ptlepmax]$.
For the sake of definiteness and consistency, in each setup we employ the central-scale NNLO+N$^3$LL $\apt$ value computed with $\mw=80.379$ GeV as our experimental proxy.
The pattern of convergence against variations of $[\ptlepmin,\ptlepmid,\ptlepmax]$ largely reflects our considerations below Eq.~\eqref{eq:Aptlep}. We also remark the need of N$^3$LL resummation for a sizeable reduction of theoretical uncertainty, and a precise $\mw$ determination.

\paragraph{Discussion.}
The asymmetry $\apt$ defined in Eq.~\eqref{eq:Aptlep} offers some interesting features, compared to a template fit of the whole $\ptlep$ distribution. First, it is defined in terms of inclusive rates integrated over relatively wide phase-space regions: this allows to obtain a fairly stable QCD prediction on the theoretical side, and an excellent statistical precision and the possibility to unfold detector effects on the experimental side.
Second, the asymmetry enables a determination of $\mw$ based on CCDY data which, upon including state-of-the-art pQCD predictions, is not dominated by the tuning of model parameters on NCDY measurements.
Third, through its linear dependence on $\mw$, the asymmetry offers the possibility to cleanly disentangle the impact on $\mw$ determination of all effects contributing to the $\ptlep$ spectrum. On top of the pQCD predictions scrutinised in this paper, which constitute a robust starting point, it is conceptually straightforward to include final-state QED radiation, as well as EW and mixed QCD-EW perturbative corrections. All of these additional effects induce modifications to $\apt$ that can be separately assessed and systematically refined. Effects of npQCD origin, relevant for a fully realistic description, can also be included as a separate component to the prediction of $\apt$, but as opposed to template-fitting, their inclusion is not instrumental for the whole $\mw$-extraction procedure.
As they involve initial-state QCD radiation, their inclusion is expected to simply induce a vertical offset to $\apt$ without altering its slope, i.e.~its sensitivity to $\mw$. This offset in turn yields a shift of the preferred $\mw$ value, which can be easily estimated thanks to the linear $\mw$-dependence of $\apt$.
The underlying npQCD model can be constrained via the simultaneous analysis of more observables, other than $\apt$: the improvement in the accuracy of this model is thus a problem fully decoupled from $\mw$ determination.

To illustrate how npQCD contributions can be consistently studied through the asymmetry $\apt$, we consider effects on $\mw$ coming from two sources: the uncertainties on the collinear proton PDFs, and those related the transfer of information from NCDY to CCDY data (further details on the results of this study can be found in the Appendix).

As for the effect of collinear PDFs, predictions for $\apt(32\, \gev,37\, \gev,47\, \gev)$ obtained using all 100 replicas of the NNPDF4.0 set
yield a PDF uncertainty of $\pm 11.5$ MeV.
More conservatively, we also consider the central replicas of the
CT18NNLO~\cite{Hou:2019efy},
MSHT20nnlo~\cite{Bailey:2020ooq}, and
NNPDF3.1~\cite{NNPDF:2017mvq} PDF sets.
The corresponding spread of $\mw$ values is of $\sim 30$ MeV.
A reduction of PDF uncertainty can be achieved by profiling PDF replicas through the simultaneous inclusion of additional information, such as data in different rapidity regions \cite{Bozzi:2015hha,Bozzi:2015zja}, all bins of the $\ptlep$ distribution \cite{Bagnaschi:2019mzi}, different $W$-boson charges at the LHC \cite{Aaboud:2017svj}.

We now discuss other effects of non-perturbative origin which affect CCDY predictions at small $\ptw$, such as the intrinsic $k_\perp$ of partons in the proton.
These npQCD effects are precisely modelled studying the $\ptz$ distribution in NCDY.
Assuming their universality \footnote{
We point out that the universality of the intrinsic-$k_\perp$ model \cite{Konychev:2005iy} can be spoiled by effects such as kinematic dependence on heavy-quark masses \cite{Berge:2005rv,Pietrulewicz:2017gxc,Bagnaschi:2018dnh}, flavour dependence \cite{Bacchetta:2018lna}, or energy-scale dependence.},
the npQCD effects can be directly applied to the CCDY simulation, inducing a shift in $\mw$.
We have investigated the interplay between the scale uncertainty of the perturbative NCDY SM description and the size of the npQCD component extracted from NCDY data (using the central NNLO+N$^3$LL NCDY prediction as pseudo-data, hence actually extracting a ``pseudo-npQCD" contribution). To this goal, we have determined one pseudo-npQCD contribution per scale choice, included it in the CCDY simulation, and assessed its impact on $\mw$ determination.
The point which emerges from this analysis is that, even if the NCDY pseudo-data are a unique set of numbers, the propagation of their information to CCDY depends on the underlying pQCD approximation, and the outcome is not unique. The CCDY results, improved with the pseudo-npQCD contribution, are spread in a range compatible with, or even larger than the scale uncertainty of the NNLO+NNLL calculation.
This result stresses the importance of using state-of-the-art pQCD results in these high-precision studies.

\paragraph{Conclusions.}
We have presented a new observable, $\apt$, sensitive to the value of the $W$-boson mass $\mw$, with promising experimental properties and robust pQCD convergence.
Its linear dependence on $\mw$ allows to systematically disentangle the impact of each contribution, perturbative or not, affecting the determination of $\mw$ and to estimate the associated uncertainty, a crucial feature for the comparison of data with SM predictions.
The study of $\apt$ highlights the importance of state-of-the-art predictions to reduce the pQCD uncertainty on $\mw$ down to the $\pm 5$ MeV level at the LHC.
We argue that, using $\apt$, an experimental error on $\mw$ at the $\pm 15$ MeV level is achievable already with Run-2 data; moreover, the possibility is given to unfold the data to particle level, easing the combination of results from different experiments.
We observe that $\apt$ can also be used in NCDY to obtain a determination of the $Z$-boson mass $m_Z$ alternative to the one based on the dilepton mass spectrum, thereby allowing a powerful cross-check of the theoretical systematics.
Given these properties, we hope that this observable will be considered for an independent determination of $\mw$ from available CCDY data.

\section*{Acknowledgements}
We thank Paolo Azzurri, Emanuele Bagnaschi, Giuseppe Bozzi, Luca Buonocore, Massimiliano Grazzini, Fabio Maltoni, Michelangelo Mangano, Pier Monni and Stefano Pozzorini for useful discussions. We thank Tobias Neumann for support in the use of \texttt{MCFM}. LR is supported by the Swiss National Science Foundation contract PZ00P2$\_$201878. PT has been partially supported by the Italian Ministry of University and Research (MUR) through grant PRIN 20172LNEEZ and by Compagnia di San Paolo through grant TORP\_S1921\_EX-POST\_21\_01. AV is supported by the Italian MUR through grant PRIN201719AVICI\_01. LR and AV thank MIAPbP for hospitality and acknowledge the Deutsche Forschungsgemeinschaft under Germany's Excellence Strategy -- EXC-2094 -- 390783311.

\appendix 
\section*{Appendix}
\begin{figure}[t]
\begin{center}
\includegraphics[width=\columnwidth]{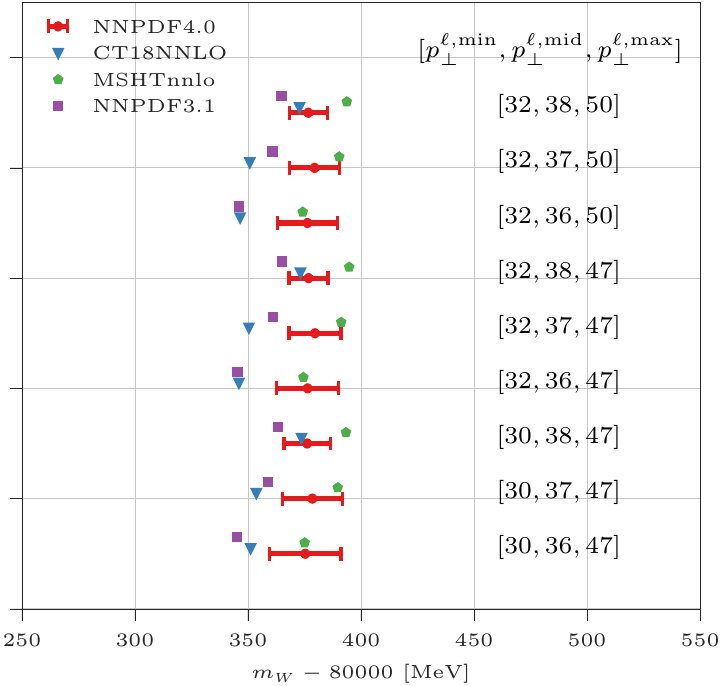}
\end{center}
\vspace{-2ex}
\caption{\label{fig:shifts-pdfs}
Same as Fig.~\ref{fig:shifts}, now comparing the range of $\mw$ values obtained with different PDF sets.
}
\end{figure}

In this Appendix we detail the study described in the main text about the impact of non-perturbative effects on $\mw$ determination. The discussion focuses on the uncertainty due to collinear PDFs, and on the modelling of an intrinsic $k_\perp$ of partons in the proton.
\paragraph{Proton-PDF uncertainties.}
Concerning the effect of different collinear PDFs, predictions for $\apt(32\, \gev,37\, \gev,47\, \gev)$ obtained with the 100 replicas of the NNPDF4.0 set yield a bundle of parallel straight lines, as expected due to the factorisation of QCD effects from $W$-boson production and decay. The intercepts with the experimental $\apt$ value yield a distribution of 100 $\mw$ values. We compute mean value and standard deviation of this distribution, obtaining at NLO+NLL with central scales a spread in $\mw$ of $\pm 11.5$ MeV.
We also consider the central replicas of the
CT18NNLO~\cite{Hou:2019efy}, MSHT20nnlo~\cite{Bailey:2020ooq}, and
NNPDF3.1~\cite{NNPDF:2017mvq} PDF sets.
The spread induced on $\mw$, using the central-scale NNLO+N$^3$LL prediction, is of $\sim 30$ MeV.
We present in Figure \ref{fig:shifts-pdfs} the results for different setups.

\begin{figure}[t]
\begin{center}
\includegraphics[width=\columnwidth]{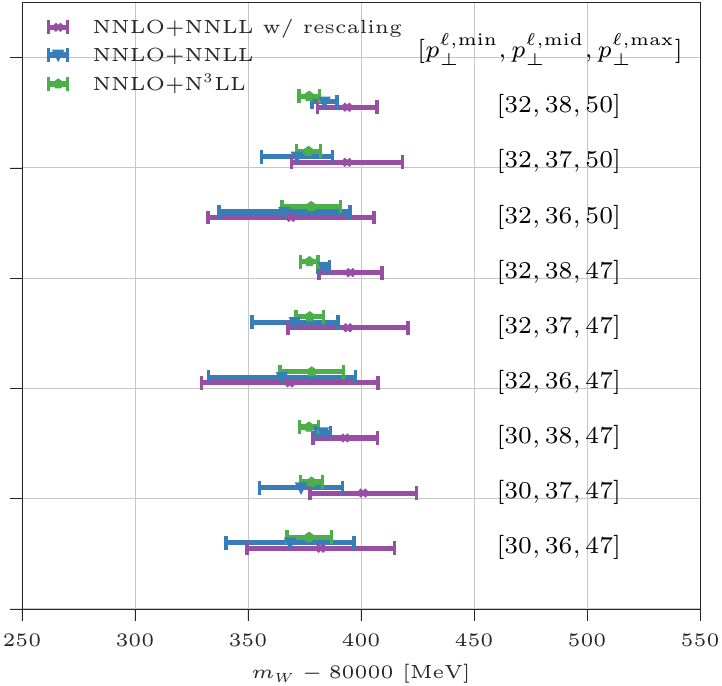}
\end{center}
\vspace{-2ex}
\caption{\label{fig:shifts-Kfact}
Same as Fig.~\ref{fig:shifts}, now including the reweighed NCDY NNLO+NNLL predictions.
}
\end{figure}

\paragraph{Modelling of the parton intrinsic $k_\perp$.}
With the following exercise, we schematically describe the encoding of information present in NCDY data and absent from a purely perturbative description of the process. We then consider the usage of such an information in the simulation of CCDY, and eventually its impact on $\mw$ determination. In particular, the pQCD stability of $\apt$ allows to study the role of scale variations in porting these effects from NCDY to CCDY.
We simulate both processes at NNLO+N$^3$LL QCD, with  $\xi_R=\xi_F=2\,\xi_Q=1$ and take the results as a proxy for experimental data (we dub them ``pseudo-data'', see also \cite{Isaacson:2022rts}).
We assume to have available an event generator with NNLO+NNLL pQCD accuracy only, and compute the NCDY $\ptz$ distribution with different scale choices.
The ratio of NNLO+NNLL predictions with pseudo-data defines a reweighing factor, as a function of $\ptz$, encoding the missing pQCD higher orders (with real data as opposed to pseudo-data it would encode npQCD effects as well). We compute one such reweighing factor per pQCD scale choice in NCDY.
We then use the NNLO+NNLL generator to simulate the CCDY process with scale variations, and reweigh all events in each variation according to their $\ptw$ value,
using the corresponding factor determined in NCDY.
Since by construction the reweighed NCDY NNLO+NNLL curves would exactly match NCDY pseudo-data, one expects to a large extent the same to happen with CCDY pseudo-data and reweighed NNLO+NNLL CCDY distributions.
We observe instead that the reweighed distributions do not exactly reproduce CCDY pseudo-data, the discrepancy being comparable with, or larger than the NNLO+NNLL scale-uncertainty band, i.e.~$\Delta\mw\sim \pm 27$ MeV from the study of $\apt(32\, \gev,37\, \gev,47\, \gev)$.
We conclude that the procedure to model npQCD effects due to an intrinsic $k_\perp$ is intertwined with the underlying pQCD formulation.
We thus expect that the same approach, using a NNLO+N$^3$LL-accurate event generator and the real data as a target, would lead to a smaller final spread in $\apt$,
providing a handle for a robust assessment of the impact of npQCD effects on the determination of $\mw$.
We present in Figure \ref{fig:shifts-Kfact} the results for different setups.

\bibliographystyle{spphys}
\bibliography{rtv}

\end{document}